\documentclass[letterpaper]{article}

\usepackage{natbib,alifeconf}  

\usepackage{algpseudocode}

\usepackage[T1]{fontenc}
\usepackage[utf8]{inputenc}
\usepackage{babel}
\usepackage{blindtext}
\usepackage{cuted}

\usepackage{subcaption}
\usepackage{graphicx}
\usepackage{float}

\title{The Effect of Noise on the Emergence of Continuous Norms and its Evolutionary Dynamics}
\author{Stavros Anagnou, Daniel Polani \and Christoph Salge \\
\mbox{}\\
Adaptive Systems Research Group, School of Computer Science, University of Hertfordshire \\
\\
s.anagnou@herts.ac.uk} 

\begin{document}
\maketitle

\begin{abstract}
We examine the effect of noise on societies of agents using an agent-based model of evolutionary norm emergence. Generally, we see that noisy societies are more selfish, smaller and discontent, and are caught in rounds of perpetual punishment preventing them from flourishing. Surprisingly, despite the detrimental effect of noise on the population, it does not seem to evolve away. We carry out further analysis and provide reasons for why this might be the case. Furthermore, we claim that our framework that evolves the noise/ambiguity of norms is a new way to model the tight/loose framework of norms, suggesting that despite ambiguous norms' detrimental effect on society, evolution does not favour clarity.
\end{abstract}

\section{Introduction \footnote{This paper has been updated to address an error in the implementation of the mutation operator, discovered after publication. We have added asterisks in the main document where the reader should refer to the errata document. }}

The social world is replete with norms, an important aspect of organising societies. Social norms reduce the degrees of freedom in the actions of individuals, making them more predictable and stabilising societies \citep{feldmanhall_resolving_2019}. Norms also enable unrelated agents to manage shared resources \citep{richerson_human_2013} , thereby extending cooperation beyond genetic relatives \citep{richerson_cultural_2016}.

Norm emergence is usually studied with discrete behaviours. Game theory tends to consider moral behaviour to be composed of discrete actions: cooperate and defect \citep{axelrod_evolutionary_1986}, hawk and dove \citep{smith_evolution_1982}, stag and hare \citep{skyrms_stag_2003}.
Other examples of discrete norms include political party affiliation, coordinating or not coordinating \citep{lewis_convention_1969, mcelreath_shared_2003} or adopting a given behaviour e.g. a possession norm \citep{epstein_growing_1996,flentge_modelling_2001}.
We know, however, that norms are not always this discrete, and a large number of norms exist on a continuous spectrum of behaviour, e.g. what amount is acceptable to take from a shared resource, how fast you walk, how close you stand next to someone during a conversation \citep{kelly_psychology_2021}. These have received much less attention in terms of modelling \citep{le_evolutionary_2007}, with the exception of continuous opinions as modelled in the closely related field of opinion dynamics \citep{flache_models_2017}.

Previous work using continuous behaviour includes continuous iterated prisoners dilemma \citep{le_evolutionary_2007} (on a scale of 0 complete defection, to 1, cooperation). 
In general, cooperation in continuous dilemmas is less stable and it is harder for cooperation to invade a population of defectors \citep{le_evolutionary_2007}. \cite{bendor_when_1991}, investigated how noise affected the success of fixed strategies in a continuous prisoners dilemma. They showed that populations of generous strategies were more successful because generous strategies avoided spiraling into rounds of mutual recrimination in noisy environments. Going beyond continuous game theory, \cite{aubert-kato_hunger_2015} investigated the emergence of frugal and greedy behaviours in an embodied version of a dilemma where agents varied in how long they exploited a food source -- the longer it exploits the food source, the more selfish the agent is. 
\cite{michaeli_norm_2015} showed how “liberal” and “conservative” punishment regimes can affect the polarisation of a continuous opinion. Further, previous work on iterated prisoners dilemma by \cite{ashlock_understanding_2006} showed that even small differences in implementation, e.g. representation choice, can lead to significantly different dynamics.

We intend to combine these previous elements together to investigate the effects of noise on the emergence of continuous social norms. We investigate this in an evolutionary agent-based simulation, comparing agent societies with deterministic and probabilistic behaviours to see if noise significantly changes the dynamics of the society, i.e. norm emergence and other properties of agent societies. 
To achieve this, we evolve three continuous norms. Uniquely, we also evolve the level of noise on each of these properties. This allows us to investigate the effect of noise on a continuous model of norm emergence, which is of use to modelers considering whether or not to include noise in their models. Further, by making the amount of noise a variable that is available to evolution, it allows us to study the evolutionary dynamics of noise.

We define criteria for norm emergence in a continuous system and show that our deterministic societies obey these criteria. We find that deterministic societies tend to be less selfish, less hypocritical and less discontented, with agents sharing resources more effectively and sanctioning each other less. In contrast, noisy societies tend to fall into perpetual punishment of each other despite the abundance of resources.

This begs the question, if noise is detrimental to the agent society, why does it not evolve away? We show that there does not seem to be an evolutionary pressure to eliminate noise and offer some reasons as to why this may be the case. Further, we suggest our model offers some insight into thinking about the evolution of loose and tight societies. The tightness-looseness framework looks at culture in terms of  the strength of their cultural norms (number and clarity) and strength of punishment when a norm is violated.  Tight cultures have stronger norms and punishments and loose cultures have more vague norms with less harsh punishments. This framework provides insights into the function of norms, with cultures tightening in response to threats, making them better at dealing with them. Our model expands the existing work by considering the noise inherent in looser societies that have more vague or ambiguous social norms \citep{gelfand_nature_2006,roos_societal_2015,pan_integrating_2021}.

\section{Model and Experiments}

\begin{figure}[h]
\begin{center}
\includegraphics[width=3.5 in, angle=0]{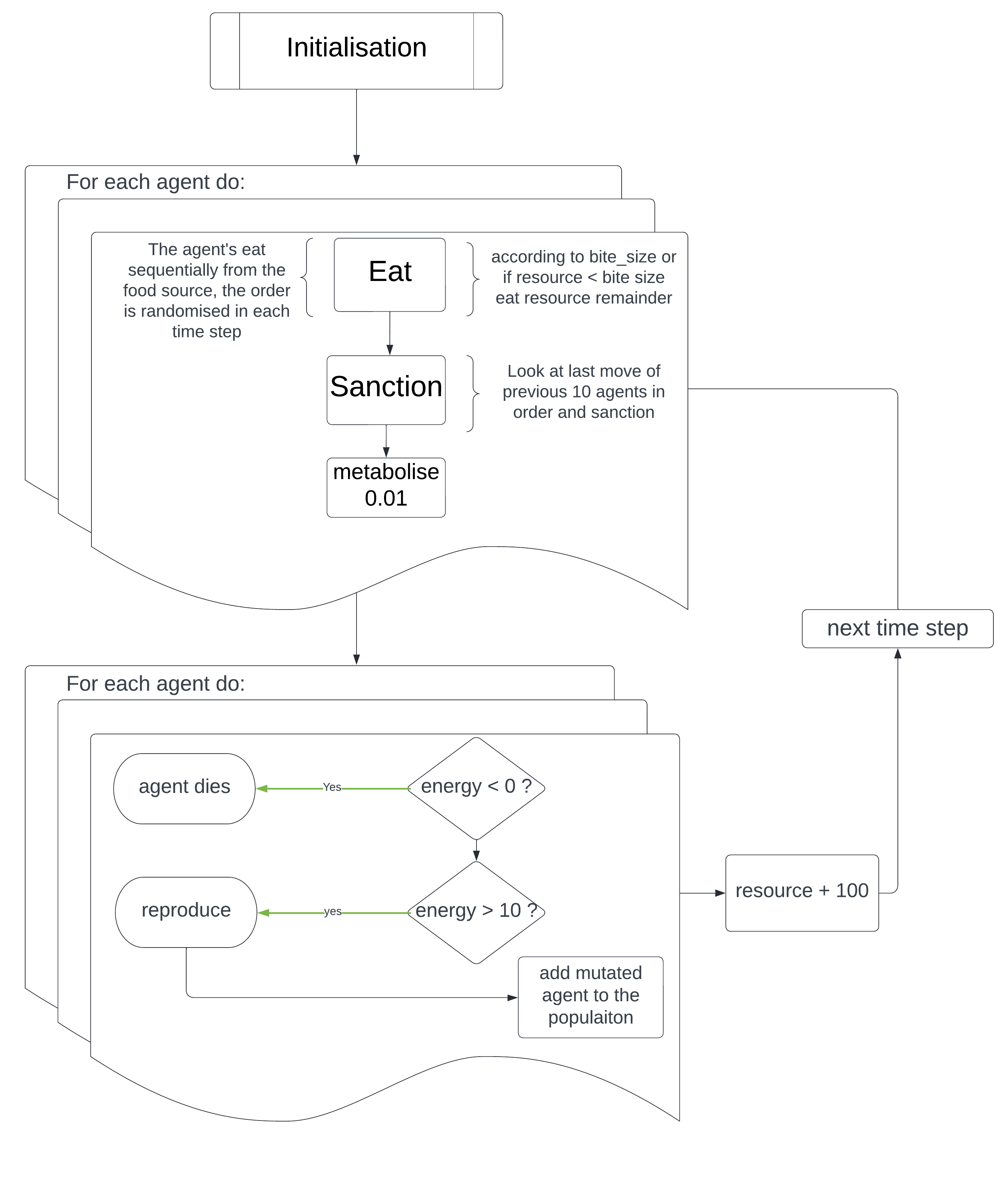}
\vskip 0.25cm
\caption{Flow diagram describing the stages of the agent-based simulation.
}
\label{Fig 1}
\end{center}
\end{figure}

The following section introduces a multi-agent model we developed to study the effect additive noise has on continuous norm emergence. We study two experimental conditions in the model. In the deterministic case, the behavior of each agent is defined by three internal, continuous variables (Bite Size (B), Sanction Threshold (T) and Sanction Strength (S)). In the probabilistic case, we add Gaussian noise to those variables each time they are used to determine behavior. The strength (variance) of this added noise is defined for each of the three internal values by another three agent-specific values, respectively - i.e. BN, TN and SN. The simulation can be separated into different steps, as visualized in Fig. 1, which are defined as follows:

\begin{figure*}[h]
\begin{center}
\includegraphics[width=6.5 in, angle=0]{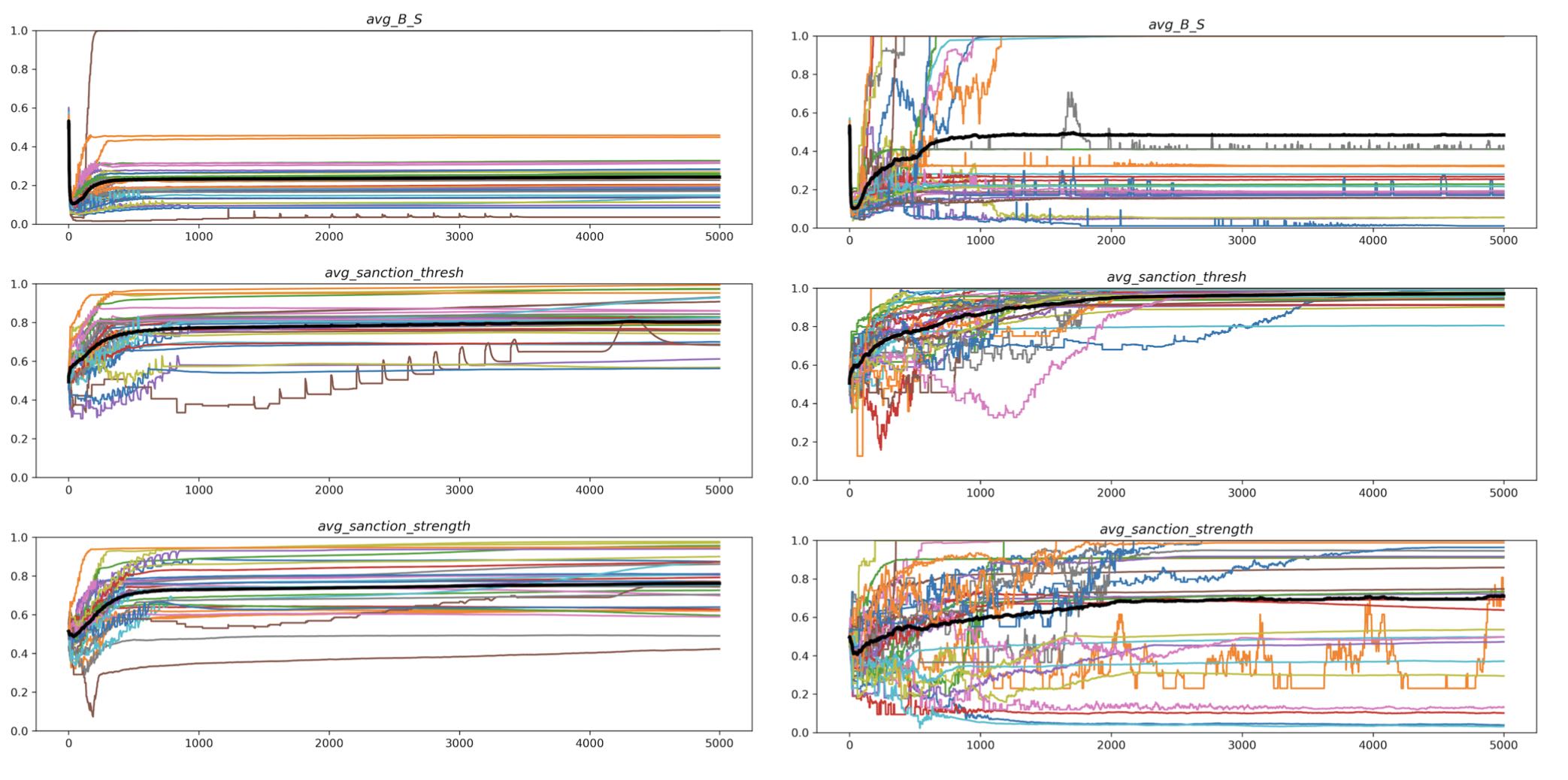}
\vskip 0.25cm
\caption{The average value of each trait in the population plotted over time. Deterministic  (left) and probabilistic (right). Individual runs are plotted as coloured lines and the average of those runs is plotted as a black line. N = 34 per condition.}
\label{Fig 1}
\end{center}
\end{figure*}

\begin{figure*}[h]
\begin{center}
\includegraphics[width=6.5 in, angle=0]{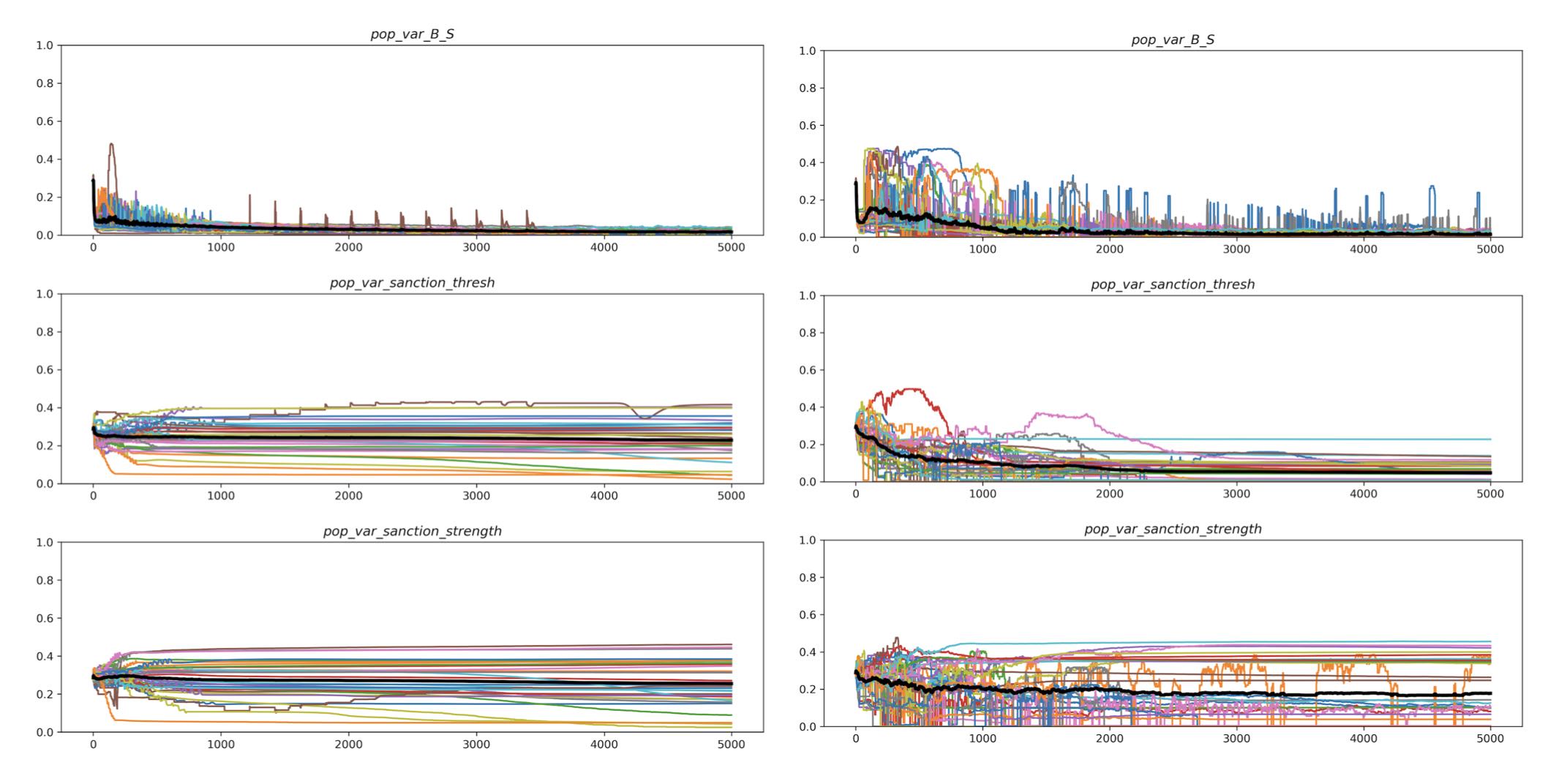}
\vskip 0.25cm
\caption{The population-level variance (the variance of a trait in each population) of each trait plotted over time. Deterministic  (left) and probabilistic (right). Individual runs are plotted as coloured lines and the average of those runs is plotted as a black line. N = 34 per condition.
}
\label{Fig 2}
\end{center}
\end{figure*}

\subsection{Initialisation}
At the beginning of the simulation, we create 100 agents and set the resource level to 1000 units. Each agent's internal values for B, S and T are initialized to uniformly random values between $0.0$ and $1.0$. For the probabilistic model, each agent's noise values (BN, TN and SN) are initialized between $0.0$ and $0.5$. Each agent's energy level is set to 10. After initialization, the simulation proceeds in rounds. Each round has a different, randomized order of all agents, and each of the following steps is performed in that order. 

\subsection{Eat} 
When it is their turn, each agent tries to consume resources according to their Bite Size (B). This value is added to their internal energy and removed from the global resource level. If there are no resources left, the agent gets no energy. The higher the value, the more greedy/selfish the agent is compared to other agents. If all agents eat at a higher Bite Size, the environment will not be able to support as many agents; thereby exhibiting 'tragedy of the commons' dynamics \citep{hardin_tragedy_1968}. 

\subsection{Sanction}
During their turn, each agent can observe the actually consumed resources of the 10 previous agents. Each agent checks if the previous agents ate more than its own internal Sanctions Threshold (T); if so, it sanctions them. In other words, T is the amount of deviance an agent tolerates before punishing another agent, i.e. what an agent finds acceptable. Sanctioning means the agent reduces the other agent's internal energy by its own Sanction Strength (S), and it also pays a sanctioning cost of 0.1 $*$ S, which is subtracted from its own energy level. 

\subsection{Metabolise}
Each agent has their energy level reduced by $0.01$ during each round.

\subsection{Death and Reproduction}
After those steps, the simulation checks if any agents have an energy level lower than 0.0, in which case they are removed from the simulation. Then any agent with an energy level larger than 10 gets to reproduce. Reproduction means we generate a copy of the agent, with the same 3 or 6 internal values,  mutated with a $0.1$ chance. [Updated]\textit{ We planned to add Gaussian noise with $0$ mean and $0.1$ variance to each of those values, but due to an error in the implementation, the value is just set to $1.0$ if a mutation is triggered. The results in this paper, are for this, extreme mutation operator. The erratum attached to the end shows comparable plots for the Gaussian mutation operator, and provides data to show that nearly all results discussed here are true for both mutation operators} [End Update] The energy of the child and parent are both set at half the parent's prior energy level. Before the next round starts, we add 100 units of resources to the resource level. 

\subsection{Creating Probabilistic Behaviour}
To create the probabalistic behaviour, we used a Gaussian function each time the agent consumed from the shared resource or punished another agent. Using their value and accompanying standard deviation (noise parameter) to create a value for the amount of food eaten B, how tolerant they are T and how harshly they punish S, e.g. for each instance of sanctions we select the strength from a normal distribution with a mean of S and a standard deviation of SN. This added noise can be interpreted either as behavioral or perceptual error.

\subsection{Implementation Choices} 
The evolutionary dynamics in this simulation are created by differential reproduction, rather than by defining an explicit fitness function and selection process. Since the order agents eat and sanction each other is sequential, it means agents at the front of the queue have an advantage, as they get to consume from the resource first. Conversely, being last in the queue also has an advantage as there is no one behind the agent to sanction them. To minimise the impact of these asymmetries on the results, we randomise this order in each round. We limit the initial values for the noise values to $[0.0,0.5]$, because some initial simulation with the full range resulted in very volatile dynamics that were hard to analyze. But after initialization, it is possible for the noise parameters to have any value between 0.0 and 1.0 so they can adapt in that direction. We explored further parameter settings not reported here (varied agent metabolism, cost of sanctions), which produced results similar to those in this paper. Note that the deterministic condition can be seen as a special case of the probabilistic condition, where the three noise parameters are very close to 0 for all agents.


\begin{figure*}[t]
\begin{center}
\includegraphics[width=6.5 in, angle=0]{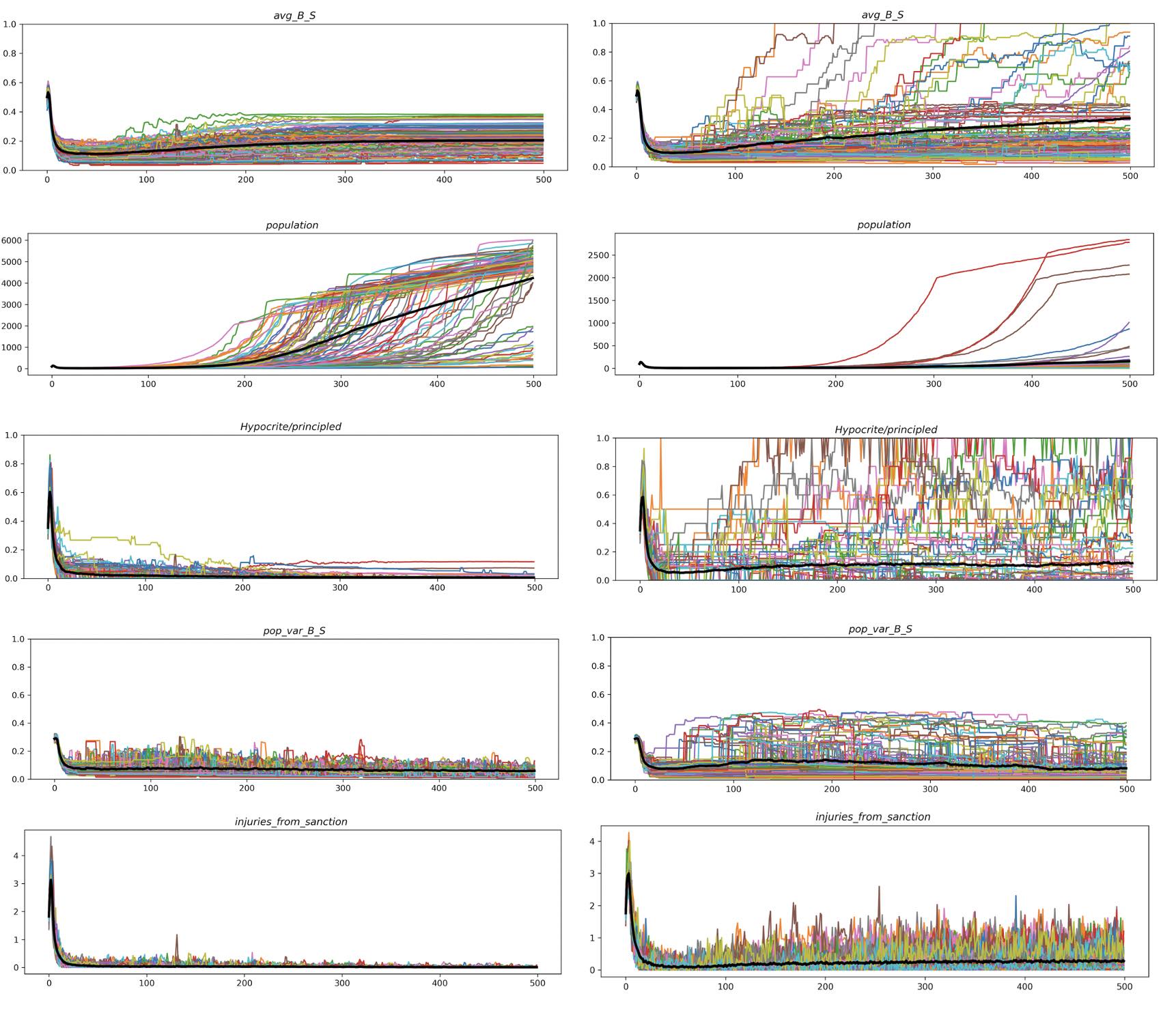}
\vskip 0.25cm
\caption{Various agent and population properties plotted over time. Deterministic (left) and probabilistic (right). Individual runs are plotted as coloured lines and the average of those runs is plotted as a black line. N = 100 per condition.}
\label{Fig 3}
\end{center}
\end{figure*}

\section{Results}

\subsection{Continuous Norm Emergence}

First we want to answer the questions, is this a model for the emergence of continuous norms? Usually in discrete models there is an arbitrary threshold, such as 80 percent of the population must possess that behaviour for it to be considered a norm (\cite{savarimuthu_norm_2011}). Since our behaviours are continuous, and its not clear what it would mean for two agents to have the same behaviour, we define criteria on how to assess norm emergence in a continuous context.
 \begin{enumerate}
       \item The behaviour converges and stabilises: Do traits decrease terms in variance across the population from where they began, and do the average behaviours stabilise? This would be indicated by a decreasing value for the variance of a given behaviour across the agent population, and a lack of change of the average behaviour over time.
       \item The behaviour the population stabilises at is arbitrary across runs to a certain extent: This criterion ensures the resultant behaviour is not fully due to environmental scaffolding, meaning when the behaviour is the only rational/viable action given the environmental constraints, e.g. a population level preference for walking over a bridge as opposed to walking across lava would not be considered a norm \citep{westra_pluralistic_2022}. This would be indicated by repeated simulations stabilizing at different average values. Note that this requires some level of randomness in the simulation, with different seeds.
       \end{enumerate}

Further to our stipulations, we clarify that we are talking about norms under the general definition of normative regularities: "A socially maintained pattern of behavioral conformity within a community."\citep{westra_pluralistic_2022}. We take this definition rather than a rule-based one, which requires higher cognitive capacities such as language (required to express the rule) \citep{kelly_psychology_2021}. Further, we think it is a wider framework that encompasses a broader range of phenomena that are of interest. This permissive of "bottom-up" approach may help us reveal the building blocks of normative cognition \citep{westra_pluralistic_2022, de_waal_towards_2010}.
In Fig. 2, we see that in our deterministic simulation, after about 200 tics, all agent traits manage to settle on a particular value. This value is arbitrary to a certain extent with different runs settling at different values. This is important as it means that the behaviour is indeed a norm and not just a product of environmental scaffolding, i.e. the only rational action given the environmental circumstances \citep{westra_pluralistic_2022}. In the probabilistic cases, it seems that norms are a lot more volatile, with average values in flux.
Further, if we look at the population level variances of each trait in Fig. 3 we see that Bite Size converges onto a much lower level than it began, thereby satisfying our criteria for norm emergence as the population converges on a shared behaviour. On the other hand, the population level variance of Sanction Threshold and Sanction Strength does not always decrease, so it is harder to make a case for the population to converge on a norm for the latter two traits.
This effect doesn't happen as much in the noisy case (Fig. 3), although there is some convergence for Sanction Threshold and Sanction Strength. However, this could be explained by the fact that probabilistic populations tend to have much lower populations than deterministic ones (Fig. 4), which could be decreasing the variance through random drift.

\subsection{Comparison Deterministic vs. Probabilistic Model}

Probabilistic agent societies are generally more selfish and have less stable cooperation*. Fig. 4 shows that deterministic societies tend to have lower Bite Sizes than probabilistic ones, meaning the societies tend to be less selfish as they are all consuming less. Further, the norm seems to be more stable in the deterministic condition, with many cases in the probabilistic condition that initially settled on low Bite Sizes breaking out into higher Bite Sizes. Particularly striking is that in 100 runs none of the deterministic runs broke away, suggesting very stable norms in the deterministic population and volatile in the noisy populations. Further, when this experiment was done with an initial noise range of $[0.0, 1.0]$ as opposed to $[0.0, 0.5]$, the average value of Bite Size went up to $0.9$, indicating very high levels of selfishness. 
Values for the other two norm traits (Sanction Threshold and Sanction Strength) are comparable between probabilistic and deterministic societies, but as with Bite Size, the runs in the probabilistic version are more volatile (Fig. 2). 
The probabilistic populations are dramatically smaller (Fig. 4), with populations of ~20 compared to thousands in the deterministic case. Only a handful of the probabilistic populations manage to reach comparable population levels to the deterministic populations.
Probabilistic societies are more hypocritical (Fig. 4). We defined hypocrisy as an agents sanction threshold being lower than its own Bite Size ($T_{self} > B_{self}$). After an initial increase at the beginning of the simulation, both deterministic and probabilistic populations see a sharp decrease. However, rates of hypocrisy are much lower in the deterministic case (around $0$) compared to hypocrites in the probabilistic case (around $0.05$), and the noisy runs are generally more volatile, with numerous runs breaking out into high numbers of hypocrites.

Initially, probabilistic and deterministic societies also have similar levels of norm convergence as measured by trait variance decreasing initially and staying at around $0.1$. However, as in Fig.~2, this level of convergence is less stable in the probabilistic condition*. Further, since the populations are much smaller in the probabilistic condition, this may bias the population level variance metric, so we should not read too much into this result. Further, the small populations mean genetic/cultural drift effects could overpower  subtle selection pressures. Despite the small populations, however, there is still a significant amount of births/deaths (data not shown here), suggesting selection is still happening. To address the effects of small populations in the future, we will have a fixed population with replacement instead of a dynamic one. But for now, since we are interested in seeing the effect on noise on the size of the populations as well, we will keep it as is.

Strangely, the reason for low populations in the probabilistic population is not a lack of resources. To see the real reason, we look at the amount of energy lost due to punishment of the agents (Fig. 4). 
This plot shows that the amount of energy was lost either as a result of damage from sanctions or cost of executing sanctions. In the deterministic case, we see a spike in sanctions at the onset if the simulation as agents punish each other due to the diversity of norms. The society then converges toward a norm as the sanctioning decreases. In the probabilistic case, it seems the societies do not adequately manage to converge most of the time, resulting in this period of adjustment never ending, leading to discontented society marred by perpetual punishment. This may be the reason agents raise their Bite Size in order to better protect themselves against punishment; in a high-noise society has hypocrites, you will likely be punished anyway, so there is no rational reason to be generous and have a low bite size. It is important here to mention our simulation is not a case of limited resources; in fact, the regrowth rate of the simulation is quite high to assess the dynamics of noise in plentiful conditions. 

In conclusion, probabilistic populations are more selfish, hypocritical, discontented and less stable while the deterministic populations managing to reach drastically higher populations with the same amount of resources.

\begin{figure}[h]
\begin{center}
\includegraphics[width=3 in, angle=0]{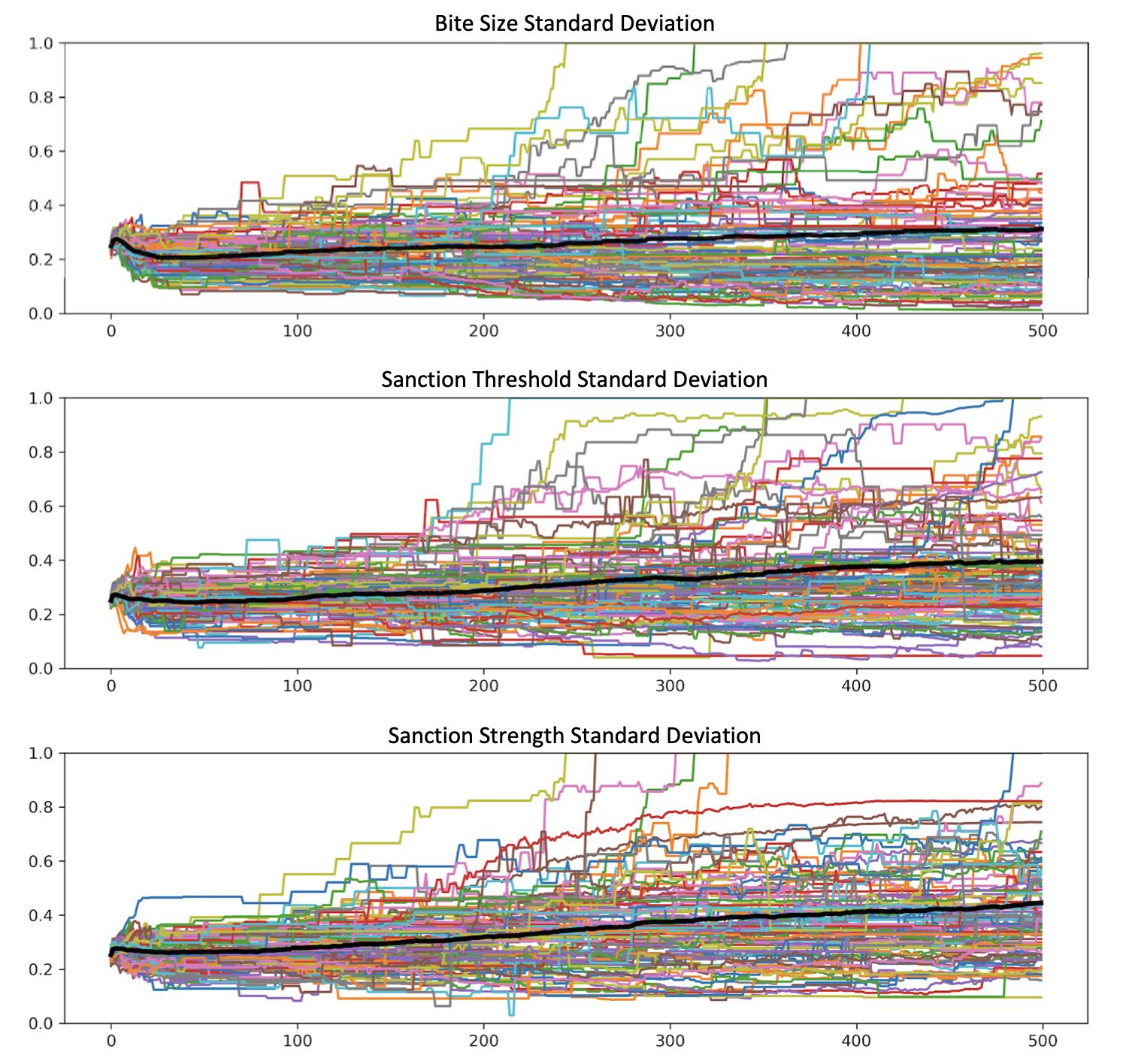}
\vskip 0.25cm
\caption{The average standard deviation (noise) for each trait plotted over time. Individual runs are plotted as coloured lines and the average of those runs is plotted as a black line. N = 100 per condition. 
}
\label{Fig 4}
\end{center}
\end{figure}

\subsection{Noise Evolution}

Given that noise in our model seems to be detrimental to a society overall, and noting that the deterministic model is just a special case of the probabilistic model, we would expect that our models evolve away the noise. But if we look at the development of the average value for the three added noise parameters, we do not see a decrease of noise (Fig. 5). On the contrary, if we just look at the first 500 steps, the average noise seems to slowly increase. This could in part be explained by a random walk, since we initialise between $[0.0,0.5]$ but allow for noise to evolve to the full range of $[0.0,1.0]$. If there was no evolutionary pressure, we would expect the noise parameters to drift towards $0.5$. 

To investigate the evolutionary dynamics of noise, we looked at how the standard deviation evolved over time (Fig. 5). We see that average noise apart from small deviations does not really seem to increase. With the individual runs looking like random walks and the average not really changing.To further investigate why this is happening and to make sure this lack of evolution is not due to the shortness of the runs, we did the following. We first ran a sample of 34 runs for each condition for 5000 time steps (10x longer than previous experiments) and then compared noisy runs that were successful (i.e. reach large populations larger than 5000, which is comparable to the deterministic case) with those that were not successful (i.e. where the population collapse or stayed at low population numbers).

\begin{figure*}[h]
\begin{center}
\includegraphics[width=6.5 in, angle=0]{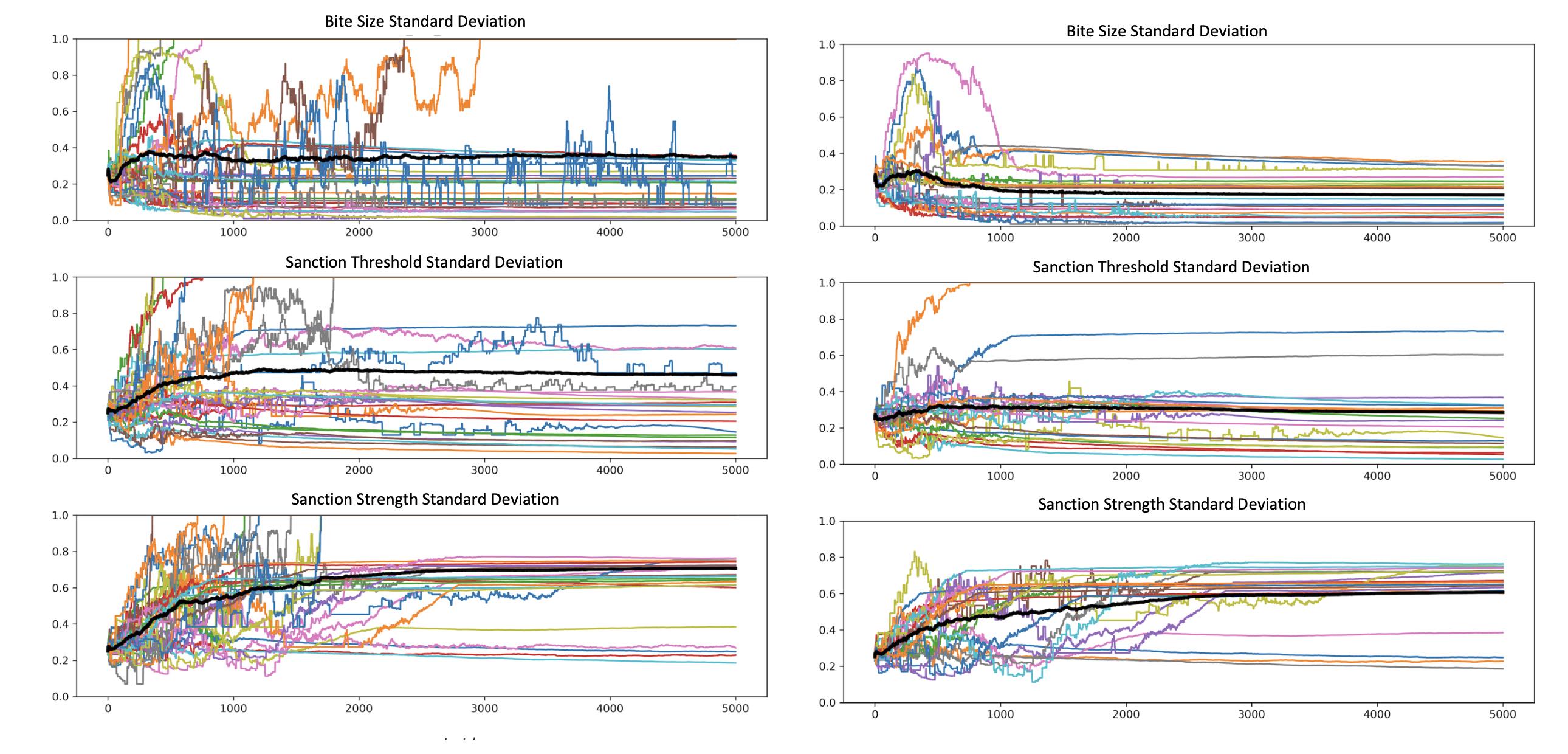}
\vskip 0.25cm
\caption{The average standard deviation (noise) for each trait plotted over time. All runs  (left) and only runs with $populations > 1000$ (right). Individual runs are plotted as coloured lines and the average of those runs is plotted as a black line. N = 34 per condition.
}
\label{Fig 5}
\end{center}
\end{figure*}

\subsection{Interpreting the Evolution of Noise}

In the longer runs, we see that noise seems to slightly increase for Bite Size for all runs Fig. 6 (left panels). But if we look at only the successful runs (right panels), it seems that on average they don’t really change in terms of their value over time. For some of the runs, this might not be so much due to evolutionary adaptation, but the fact that the runs started with low noise to begin with and happened to be better at surviving. This being said, a minor proportion of successful runs had an initial increase in noise but eventually settled at low noise values. It seems then that for Bite Size, successful runs (those that reach higher populations) are runs that, by chance, started at the low noise levels. 

If we look at the noise for Sanction Threshold (tolerance) (TN), we see a similar pattern: a slight increase in noise overall if we look at all of the runs but the majority of the successful runs start at low noise and remain there. So although the graphs show that we could reduce noise by selecting populations by their overall performance, noise level does not seem to decrease when selection occurs on a per agent basis.

In contrast to the other two traits, Sanction Strength noise (SN)  seems to increase in both successful and unsuccessful runs*, in spite of the fact that noise is initialised at [0, 0.5] (Fig. 6). Specifically, there seems to be a region between 0.6 and 0.8 where the most successful runs seem to settle, with unsuccessful runs above and below this region. One could interpret this as being a selection pressure favouring an intermediate level of Sanction Strength noise, however more experiments with runs starting at a wider range of noise would be needed to conform this.

Taken together, this results seem to suggest that although noise makes agent populations less successful, there doesn't seem to be an evolutionary process that reduces noise, in fact, in the case of Sanction Strength, evolution may increase noise.

\section{Discussion}

We present a model of continuous norm emergence to show the effects of noise on the dynamics of norm emergence and the simulation more generally.

We first showed that deterministic societies satisfy our criteria for norm emergence:
\begin{itemize}
\item 1: That they converge on behaviours compared to the start of the simulation; and 
\item 2: That the norm is somewhat arbitrary, meaning not the only rational action given the circumstances i.e. environmental scaffolding \citep{westra_pluralistic_2022}. 
\end{itemize}
Second, we showed that deterministic societies compared to noisy ones are: more able to settle on norms, distribute resources effectively (altruism), less hypocritical, less discontent and are more stable in these properties over time. In contrast, noisy societies do not prosper because they have high levels of sanctioning not seen in deterministic societies, falling into rounds of perpetual punishment. 
This result goes against the "mad man theory" hypothesis, where an adversary in a negotiation is more likely to stand down if they think their opponent (in our simulation, the agent sanctioning) is unpredictable to avoid provoking them (\cite{mcmanus_revisiting_2019}). In our simulation, it appears that noisy punishment only seems to incentivise agents to increase their provoking behavior (breaking norms by increasing Bite Size). This occurs because they will be punished whether they obey a norm or not, so they might as well consume as much as they can to increase their chances of survival. Although, the detrimental effects of noisy sanctions may be due to the fact that both the punishment behaviours (S and T) and eating behaviour (B) are noisy, further analysis by adding noise to only one of the behaviours e.g. B or T would be needed to confirm this. 

The widespread instability of noisy societies raised the question: if noise is detrimental, wouldn't it evolve away? Through further analysis, we showed that although populations with low levels of noise ended up being more successful (reaching higher populations), there wasn't an evolutionary trend toward reducing noise, in fact in some cases there seemed to be an evolutionary pressure to increase noise.

We offer some explanations why this may be happening. Firstly, for noise to be selected against there might need be group selection, as deterministic societies tend to be much larger, they would be able to out compete smaller noisy groups. Secondly, although high levels of noise are detrimental to the group, there might be a benefit to the individual of having noise; perhaps it enables them to avoid punishments. Further, one agent lowering their noise would not necessarily benefit them enough to dominate the population if the rest of society still has high levels of noise. 

A further contribution of this paper is that it may offer a way to model/think about the evolution of cultural tightness/looseness \citep{gelfand_nature_2006}. Which is defined as 1. strength of sanctioning (tolerance to deviance from norms) and 2. strength of social norms (number and clarity).  Tight cultures have stronger norms and punishments and loose cultures have vaguer norms with less harsh punishments \citep{gelfand_nature_2006, roos_societal_2015}. We refine 1. by differentiating between tolerance to deviation (i.e. sanction threshold) and also the strength of punishment when someone deviates (i.e. sanction strength). Further, our model could be a new way to model the evolution of clarity of social norms using noise (2). Finally,  computer models studying tightness and looseness assume discrete behaviours, we relax this assumption by grounding our study in a continuous modelling framework \citep{pan_integrating_2021}. Taking this lens on our simulation, we could claim that the clarity of norms (tightness) can't be evolved despite there being an ostensible selection pressure against it. This is a peculiar finding, as it would imply although societies with vague rules are at a disadvantage, ambiguity persists.

\subsection{Future Work}

In our future work we would like add the following extensions. Currently the regrowth rate of the shared resource is static. We could vary this and make the resource  growth dynamic by having "seasons" and see the resultant dynamics. Further, we only studied vertical cultural transmission but didn’t include horizontal transmission, where individuals in the same generation copy each other's strategies. In contrast to other models of tightness and looseness, our model has three norms instead of one, we could add more norms and analyse the interplay between the "tightness" and "looseness" of different norms. Finally, to further study the evolutionary dynamics of noise; we should compare different combinations of traits with and without noise, e.g. have a noisy punishment threshold but deterministic bite size and punishment strength. 

\section{Acknowledgements}
We would like to thank Niki Papadogiannaki, Imran Khan and the anonymous reviewers for their helpful comments and feedback. Stavros Anagnou is supported by a studentship from the University of Hertfordshire.

\footnotesize
\bibliographystyle{apalike}
\bibliography{CITATIONSFORALIFE} 

\begin{thebibliography}{}

\bibitem[Ashlock et~al., 2006]{ashlock_understanding_2006}
Ashlock, D., {Eun-Youn Kim}, and Leahy, N. (2006).
\newblock Understanding representational sensitivity in the iterated prisoner's
  dilemma with fingerprints.
\newblock {\em IEEE Transactions on Systems, Man and Cybernetics, Part C
  (Applications and Reviews)}, 36(4):464--475.

\bibitem[Aubert-Kato et~al., 2015]{aubert-kato_hunger_2015}
Aubert-Kato, N., Witkowski, O., and Ikegami, T. (2015).
\newblock The {Hunger} {Games}: {Embodied} agents evolving foraging strategies
  on the frugal-greedy spectrum.
\newblock In {\em {ECAL} 2015}, pages 357--364. The MIT Press.

\bibitem[Axelrod, 1986]{axelrod_evolutionary_1986}
Axelrod, R. (1986).
\newblock An {Evolutionary} {Approach} to {Norms}.
\newblock {\em American Political Science Review}, 80(4):1095--1111.

\bibitem[Bendor et~al., 1991]{bendor_when_1991}
Bendor, J., Kramer, R.~M., and Stout, S. (1991).
\newblock When in {Doubt}...: {Cooperation} in a {Noisy} {Prisoner}'s
  {Dilemma}.
\newblock {\em Journal of Conflict Resolution}, 35(4):691--719.

\bibitem[de~Waal and Ferrari, 2010]{de_waal_towards_2010}
de~Waal, F.~B. and Ferrari, P.~F. (2010).
\newblock Towards a bottom-up perspective on animal and human cognition.
\newblock {\em Trends in Cognitive Sciences}, 14(5):201--207.
\newblock Number: 5.

\bibitem[Epstein and Axtell, 1996]{epstein_growing_1996}
Epstein, J.~M. and Axtell, R. (1996).
\newblock {\em Growing artificial societies: social science from the bottom
  up}.
\newblock Complex adaptive systems. Brookings Institution Press, Washington,
  D.C.

\bibitem[FeldmanHall and Shenhav, 2019]{feldmanhall_resolving_2019}
FeldmanHall, O. and Shenhav, A. (2019).
\newblock Resolving uncertainty in a social world.
\newblock {\em Nature Human Behaviour}, 3(5):426--435.
\newblock Number: 5.

\bibitem[Flache et~al., 2017]{flache_models_2017}
Flache, A., Mäs, M., Feliciani, T., Chattoe-Brown, E., Deffuant, G., Huet, S.,
  and Lorenz, J. (2017).
\newblock Models of {Social} {Influence}: {Towards} the {Next} {Frontiers}.
\newblock {\em Journal of Artificial Societies and Social Simulation}, 20(4):2.

\bibitem[Flentge et~al., 2001]{flentge_modelling_2001}
Flentge, F., Polani, D., and Uthmann, T. (2001).
\newblock Modelling the {Emergence} of {Possession} {Norms} using {Memes}.
\newblock {\em Journal of Artificial Societies and Social Simulation}, vol.
  4(no. 4).

\bibitem[Gelfand et~al., 2006]{gelfand_nature_2006}
Gelfand, M.~J., Nishii, L.~H., and Raver, J.~L. (2006).
\newblock On the nature and importance of cultural tightness-looseness.
\newblock {\em Journal of Applied Psychology}, 91(6):1225--1244.

\bibitem[Hardin, 1968]{hardin_tragedy_1968}
Hardin, G. (1968).
\newblock The {Tragedy} of the {Commons}: {The} population problem has no
  technical solution; it requires a fundamental extension in morality.
\newblock {\em Science}, 162(3859):1243--1248.

\bibitem[Kelly and Setman, 2021]{kelly_psychology_2021}
Kelly, D. and Setman, S. (2021).
\newblock The {Psychology} of {Normative} {Cognition}.
\newblock In {\em The {Stanford} {Encyclopedia} of {Philosophy}}. Metaphysics
  Research Lab, Stanford University, spring 2021 edition.

\bibitem[Le and Boyd, 2007]{le_evolutionary_2007}
Le, S. and Boyd, R. (2007).
\newblock Evolutionary dynamics of the continuous iterated {Prisoner}'s
  dilemma.
\newblock {\em Journal of Theoretical Biology}, 245(2):258--267.

\bibitem[Lewis, 1969]{lewis_convention_1969}
Lewis, D.~K. (1969).
\newblock {\em Convention: a philosophical study}.
\newblock Blackwell, Oxford, nachdr. edition.

\bibitem[Mathew et~al., 2013]{richerson_human_2013}
Mathew, S., Richerson, P.~J., and Van~Veelen, M. (2013).
\newblock Human {Cooperation} among {Kin} and {Close} {Associates} {May}
  {Require} {Enforcement} of {Norms} by {Third} {Parties}.
\newblock In {\em Cultural {Evolution}}. The MIT Press.

\bibitem[McElreath et~al., 2003]{mcelreath_shared_2003}
McElreath, R., Boyd, R., and Richerson, P. (2003).
\newblock Shared {Norms} and the {Evolution} of {Ethnic} {Markers}.
\newblock {\em Current Anthropology}, 44(1):122--130.
\newblock Number: 1.

\bibitem[McManus, 2019]{mcmanus_revisiting_2019}
McManus, R.~W. (2019).
\newblock Revisiting the {Madman} {Theory}: {Evaluating} the {Impact} of
  {Different} {Forms} of {Perceived} {Madness} in {Coercive} {Bargaining}.
\newblock {\em Security Studies}, 28(5):976--1009.

\bibitem[Michaeli and Spiro, 2015]{michaeli_norm_2015}
Michaeli, M. and Spiro, D. (2015).
\newblock Norm conformity across societies.
\newblock {\em Journal of Public Economics}, 132:51--65.

\bibitem[Pan et~al., 2021]{pan_integrating_2021}
Pan, X., Gelfand, M., and Nau, D. (2021).
\newblock Integrating evolutionary game theory and cross-cultural psychology to
  understand cultural dynamics.
\newblock {\em American Psychologist}, 76(6):1054--1066.

\bibitem[Richerson et~al., 2016]{richerson_cultural_2016}
Richerson, P., Baldini, R., Bell, A.~V., Demps, K., Frost, K., Hillis, V.,
  Mathew, S., Newton, E.~K., Naar, N., Newson, L., Ross, C., Smaldino, P.~E.,
  Waring, T.~M., and Zefferman, M. (2016).
\newblock Cultural group selection plays an essential role in explaining human
  cooperation: {A} sketch of the evidence.
\newblock {\em Behavioral and Brain Sciences}, 39:e30.

\bibitem[Roos et~al., 2015]{roos_societal_2015}
Roos, P., Gelfand, M., Nau, D., and Lun, J. (2015).
\newblock Societal threat and cultural variation in the strength of social
  norms: {An} evolutionary basis.
\newblock {\em Organizational Behavior and Human Decision Processes},
  129:14--23.

\bibitem[Savarimuthu and Cranefield, 2011]{savarimuthu_norm_2011}
Savarimuthu, B. T.~R. and Cranefield, S. (2011).
\newblock Norm creation, spreading and emergence: {A} survey of simulation
  models of norms in multi-agent systems.
\newblock {\em Multiagent and Grid Systems}, 7(1):21--54.

\bibitem[Skyrms, 2003]{skyrms_stag_2003}
Skyrms, B. (2003).
\newblock {\em The {Stag} {Hunt} and the {Evolution} of {Social} {Structure}}.
\newblock Cambridge University Press, 1 edition.

\bibitem[Smith, 1982]{smith_evolution_1982}
Smith, J.~M. (1982).
\newblock {\em Evolution and the {Theory} of {Games}}.
\newblock Cambridge University Press, 1 edition.

\bibitem[Westra and Andrews, 2022]{westra_pluralistic_2022}
Westra, E. and Andrews, K. (2022).
\newblock A pluralistic framework for the psychology of norms.
\newblock {\em Biology \& Philosophy}, 37(5):40.

\end{thebibliography}

\clearpage

\section{Erratum}

This is an erratum to \textbf{The Effect of Noise on the Emergence of Continuous Norms and its Evolutionary Dynamics}, published in the Proceedings of the 2023 Conference on Artificial Life. After publication, we discovered an error in the mutation operator. In this correction, we will describe the error, outline what dynamics would produce the results we initially presented, and then compare them with simulation results with the mutation operator working as described in the original paper. We go through the claims of the original paper, discussing any possible differences in the outcomes. Overall, the main results of the paper still hold, although with some minor differences.

\subsection{Error description}
The faulty mutation operator in the original paper would, for each gene in the genome, check if it would mutate (vs. the mutation probability), but then, instead of adding an amount of Gaussian noise to mutate the variable, would set it to 1.0. This was caused by the code that was supposed to limit the variable to a range between 0.0 and 1.0. The variable limits were switched during the function call, so the code applied a lower limit of 1.0 to all values after mutation. While this still implements ``a'' form of mutation, these were not the dynamics intended by us, nor the ones described in the paper. The plots in the original paper were produced with the mutation operator working like this in all instances mentioned. In the following section, we will provide updated plots with the mutation operator working as described in the original paper.


\subsection{Norm emergence (Results Hold)}
In our amended simulation norms still emerge according to our definition: They converge on a particular value through a reduction in variance (Fig. 8) are to a certain extent arbitrary (Fig. 7) (path dependent), i.e. they are not determined by environmental scaffolding. Therefore our definition is intact.

When comparing deterministic and probabilistic norm emergence we also see largely similar dynamics to the original simulation; reduction in variance in the bite size norm (Fig. 8) in both cases but the average is more chaotic in the probabilistic case i.e. unstable norm emergence (Fig. 7). In addition, we also ran the simulations for longer as well and showed that bite size (in both  deterministic and probabilistic conditions) seems to slowly creep upwards over time. This change is probably due to the fact of the evolutionary incentive to consume more resource in order to reproduce. Ordinarily, this would trigger sanctions and prevent an increase in bite size in  the population; however, these changes are smaller than the variance in the population, therefore not triggering significant sanctions to counteract them. This effect is due to the continuous nature of the simulation.

N.B. Smaller populations still have an issue that there is less variance so may be harder to make that point for norm convergence in the probabilistic case.

\begin{figure*}[ht]
 \centering
    \begin{subfigure}{0.49\textwidth}
    \centering
        \includegraphics[width=\textwidth]{FIG_1.jpg}
        \caption{This is the old data with the malfunctioning mutation operator, \\$N = 34$ per condition. Each simulation is 5000 time steps.}
        \label{fig:first}
    \end{subfigure}
    \hfill
    \begin{subfigure}{0.49\textwidth}
    \centering
        \includegraphics[width=\textwidth]{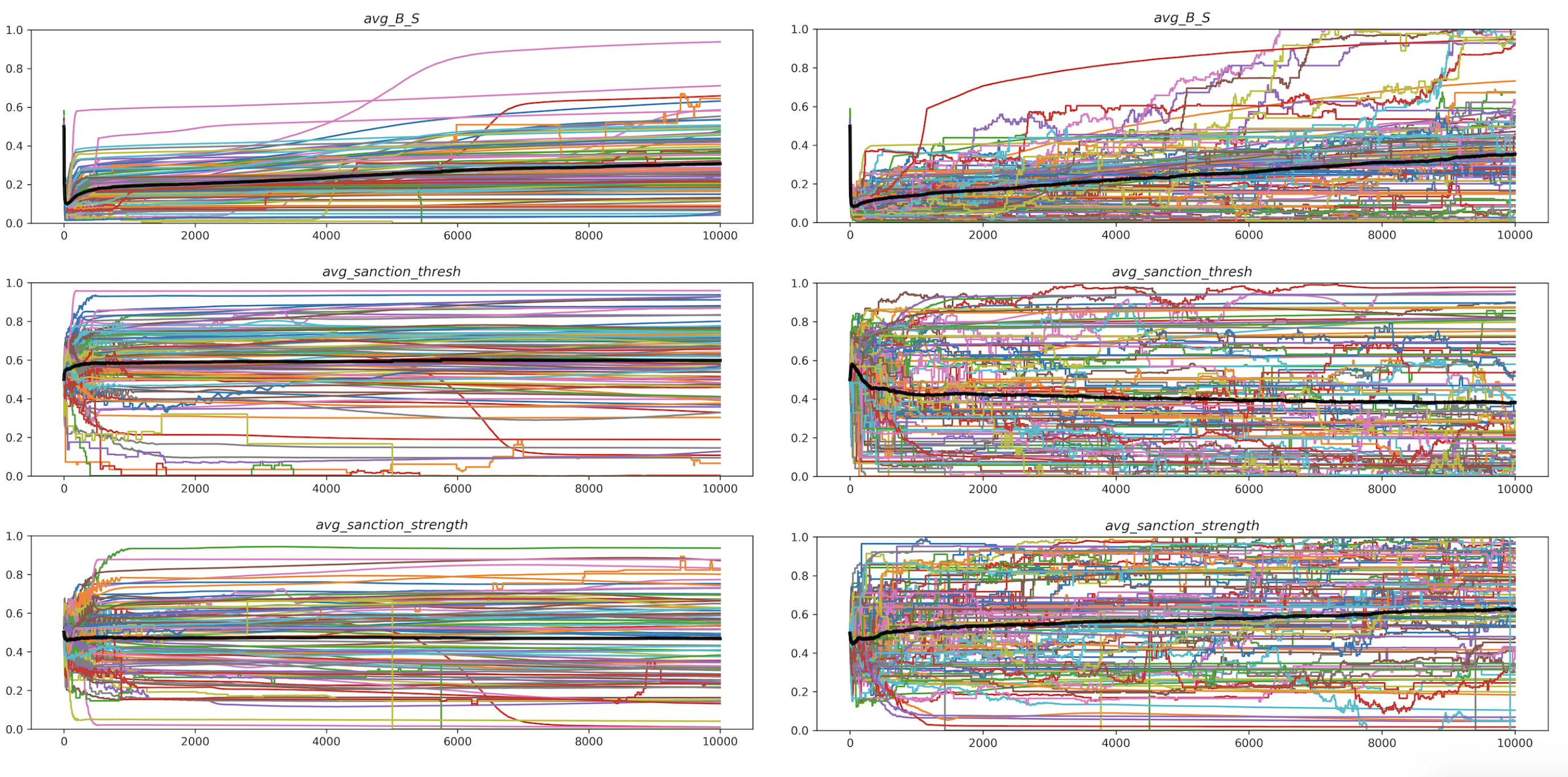}
        \caption{This is the new data with the functioning mutation operator, \\$N = 100$ per condition. Each simulation is 10000 time steps.}
        \label{fig:second}
    \end{subfigure}     
\caption{The average value of each trait in the population plotted over time. Deterministic (left) and probabilistic (right). Individual runs are plotted as coloured lines and the average of those runs is plotted as a black line. The main trends carry over from the old simulation to the new one: both show that the deterministic simulations seem to settle on a norm whereas the probabilistic simulations are a lot more chaotic. There are some differences 1: that average bite size is similar between conditions in the new simulation but not on the old. 2: the sanction threshold and sanction strength averages tend toward the middle instead of toward the top of the range.}
\label{fig:figures}
\end{figure*}

\subsection{Probabilistic vs Deterministic Population Level Properties}

\begin{enumerate}
       
       \item Probabilistic populations are smaller, this result carries over to the new simulation (result holds). A difference here is that that the population average is larger for both conditions in the new simulation, however this is due to the simulation being run for more time steps in the new simulations so the simulations grow more than in the old simulation.
       
       \item Probabilistic populations are more hypocritical, this result carries over to the new simulation (result holds/is stronger). 
      
       \item Probabilistic populations have less variance but this may be due to smaller populations in the probabilistic case (conclusion unclear). This is in contrast to the original paper result where probabilistic populations had more variance despite a smaller population.
       
       \item Probabilistic agents seem to punish each other more. In fact this difference not only remains in the new data but is even larger in the new data (results hold/is stronger).
       
       \item Bite sizes are largely the same across both populations (result no longer holds). This contrasts with the original result of the probabilistic population being more selfish (higher average bite-size).  This being said, despite having a much smaller population compared the the deterministic population (and therefore less potential competition) the probabilistic bite size size is similar. So bite sizes may be similar but only due to the difference in population size. (Fig. 8 b, top row).
       \end{enumerate}

\begin{figure*}
\centering
\begin{subfigure}{0.49\textwidth}
    \includegraphics[width=\textwidth]{FIG_3.jpg}
    \caption{This is the old data with the malfunctioning mutation operator,each simulation was run for 500 time steps.}
    \label{fig:first}
\end{subfigure}
\hfill
\begin{subfigure}{0.49\textwidth}
    \includegraphics[width=\textwidth]{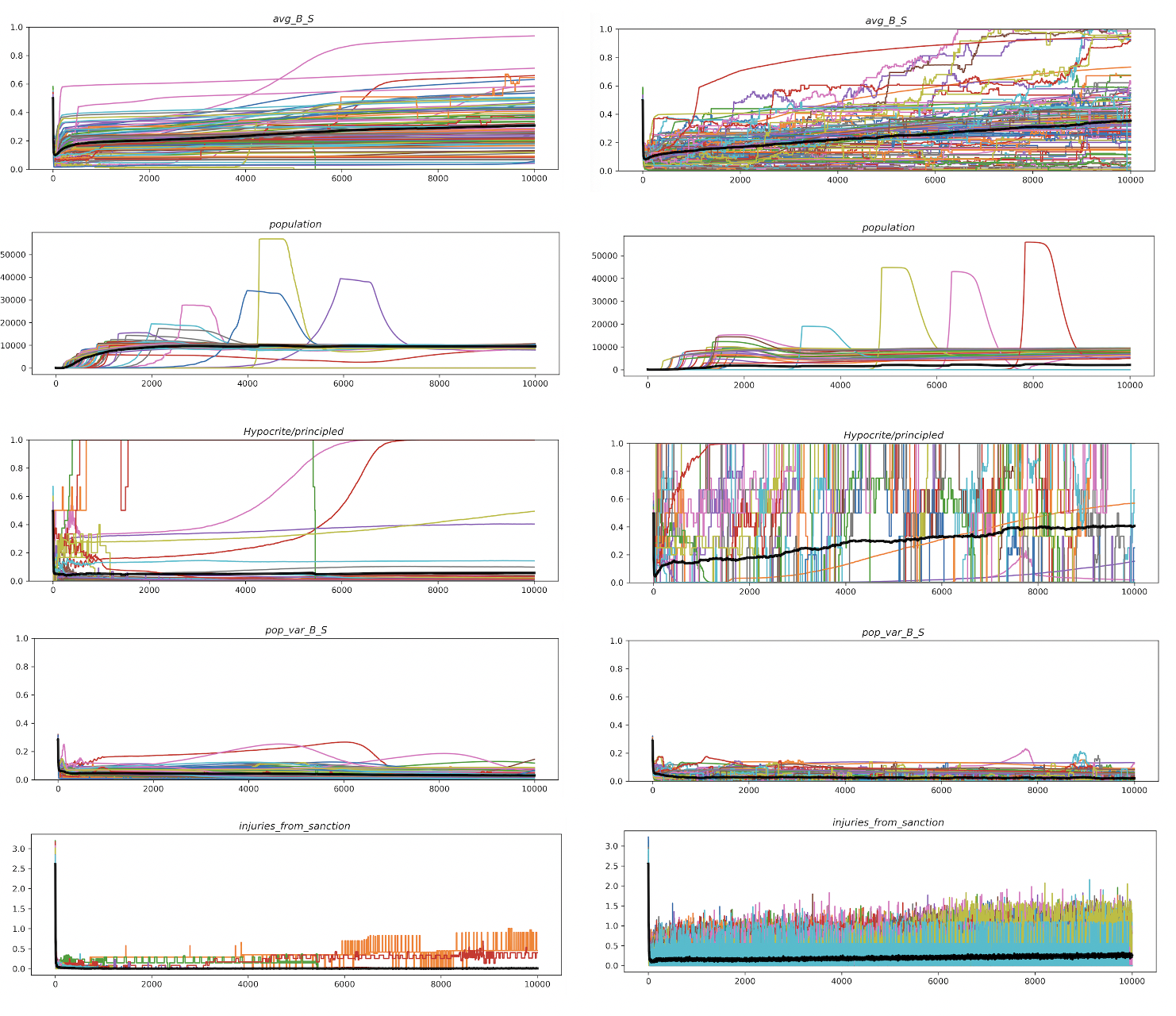}
    \caption{This is the new data with the functioning mutation operator, each simulation was run for 10000 time steps.}
    \label{fig:second}
\end{subfigure}        
\caption{Various agent and population properties plotted over time. Deterministic (left) and probabilistic (right). Individual runs are plotted as coloured lines and the average of those runs is plotted as a black line. N = 100 per condition. Although most trends are preserved, there are a few differences: 1. the bite size average in the original population is different when comparing probabilistic and deterministic conditions, with the average bite size tending to be more selfish. In the new simulation there is no such difference. 2. The population differences remain (deterministic being larger) although the average is smaller in the new simulation. 3. There are more hypocrites in the probabilistic conditions in the new simulation, confirming the old simulation. 4. The amount of variance is larger for probabilistic in the old simulation, but this is not the case in the new simulation, with the variance being the same for both probabilistic and deterministic. Finally, under the "injuries due to sanctions" plot (shows energy agents lose due to punishment) the new simulation confirms the old simulation results; with probabilistic populations losing more energy due to higher levels of punishment.}
\label{fig:figures}
\end{figure*}

\begin{figure*}
\centering
\begin{subfigure}{0.49\textwidth}
    \includegraphics[width=\textwidth]{FIG_5.jpg}
    \caption{This is the old data with the malfunctioning mutation operator. N = 34 per condition. Simulation was 5000 steps}
    \label{fig:first}
\end{subfigure}%
\hfill
\begin{subfigure}{0.49\textwidth}
    \includegraphics[width=\textwidth]{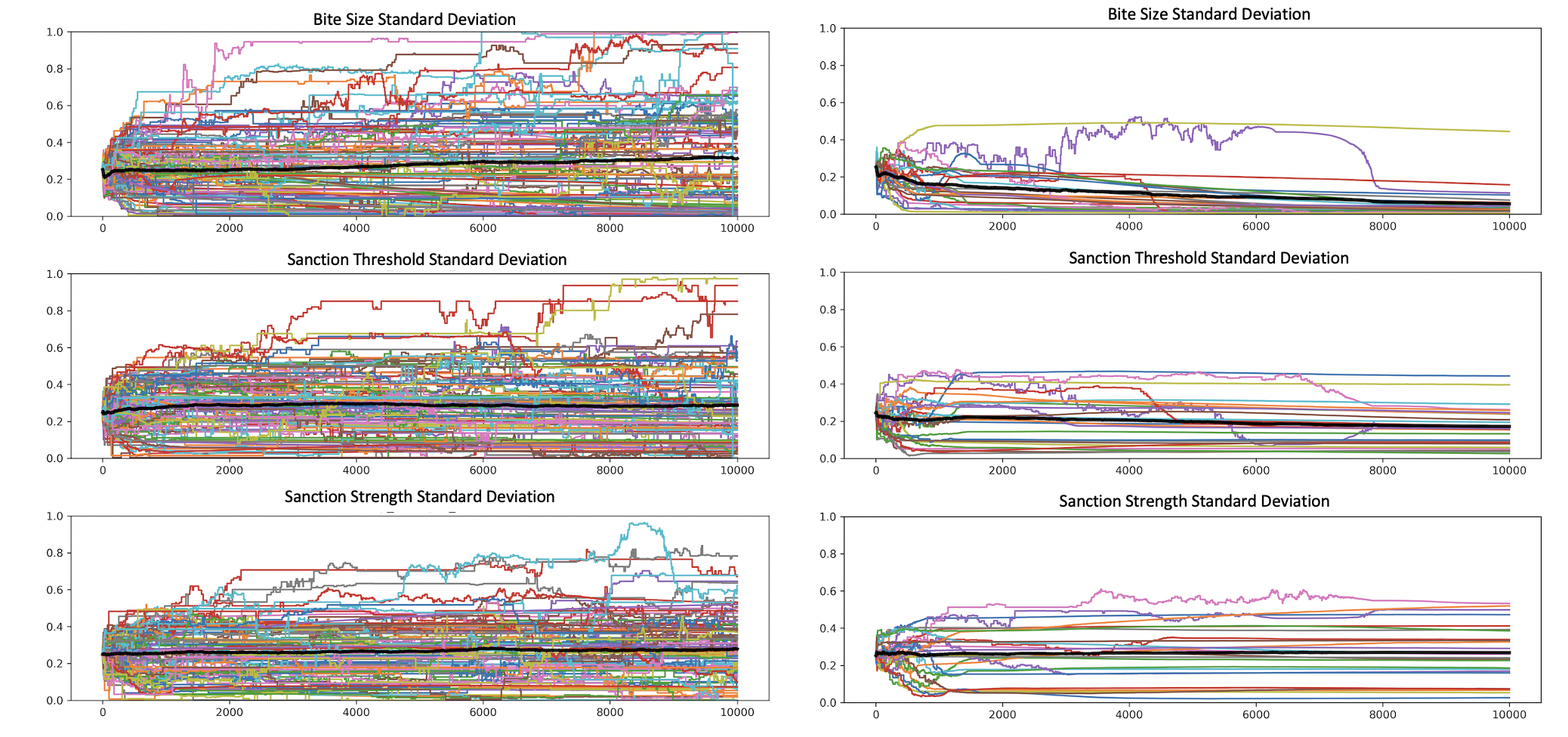}
    \caption{This is the new data with the functioning mutation operator. N = 100 per condition. Simulation was 10000 steps}
    \label{fig:second}
\end{subfigure}  
\caption{The average standard deviation (noise) for each trait plotted over time. All runs  (left) and only runs with $populations > 1000$ (right). Individual runs are plotted as coloured lines and the average of those runs is plotted as a black line. In the old simulation we see that there is generally a small increase in noise when all runs are considered and noise stays the same when only successful runs are considered (with the exception of sanction strength noise which increases). This is a strange overall result in where despite the negative effects of noise, it doesn't evolve away.
In the new simulation, the main trend is preserved in that noise doesn't evolve away for any of the traits, however it stays the same instead of increasing. Further, it seems that if we look at only successful runs (This is when select runs at the group level for their population) bite size noise seems to decrease, further confirming that noise is detrimental at the group level and yet isn't selected for at the individual level. Overall, a lower level of bite size noise is beneficial to the group, strangely, it is not selected against by evolution.}
\label{fig:figures}
\end{figure*}

\subsection{Evolution of Noise (Results Hold)}

Regarding the evolutionary dynamics of noise, the main claim holds: noise does not seem to evolve away despite being detrimental to the population. However, noise does not increase in some cases as in the original results (see sanction threshold noise in Fig 9a).  In the case of the new mutation operator, it seems as if noise just does not evolve, suggesting that despite noise being detrimental, there may be no local gradient for it to evolve away. We also analysed the “successful” runs (where we show only the runs that have populations > 1000). We see that when we select runs at the group level for their population size, sanction noise remains at the same level. However, bite-size noise seems to decrease over time, suggesting that for runs that we artificially select for at the group level, there is a decrease in noise. However, it should be noted that these are the subset of runs that decreased in noise via evolutionary drift, which we then selected for and plotted after the simulation. These changes do not occur due to individual-level selection alone (Fig. 9b). Overall, these results suggest that although lower noise is better for group success (population size) overall, individual-level selection is not enough for noise to evolve away; this is the same conclusion as with the original simulation.

\end{document}